\documentclass[]{elsart}
\usepackage{graphicx,natbib,amssymb,psfig}
\journal{New Astronomy}
\def\astrobj#1{#1}
\begin{document}
\begin{frontmatter}
\title{On the rotation periods of the components of the triple system TYC\,9300-0891-1AB/TYC\,9300-0529-1 in the Octans Association}
\author[INAF]{Sergio Messina\corauthref{cor}},
\corauth[cor]{Corresponding author.}
\ead{sergio.messina@oact.inaf.it}
\author[KKO]{Berto Monard},
\author[SAAO]{Hannah L. Worters},
\author[CFA]{Gordon E. Bromage},
\author[INAF]{Richardo Zanmar Sanchez}
\address[INAF]{INAF- Catania Astrophysical Observatory, via S.Sofia, 78 I-95123 Catania, Italy} 
\address[KKO]{Klein Karoo Observatory,   Western Cape, South Africa}
\address[SAAO]{South African Astronomical Observatory, P.O. Box 9, Observatory, 7935, South Africa}
\address[CFA]{Centre for Astrophysics, University of Central Lancashire, Presto PR1 2HE, UK}

\begin{abstract} 
Stellar rotation depends on different parameters such as age, mass, initial chemical composition, initial angular momentum, and environment characteristics. The range of values of these parameters causes the dispersion in the rotation period distributions observed in young stellar clusters/associations. We focus our investigation on the effects of different circumstellar environments on stellar rotation. More specifically, we consider the effects of a perturber stellar companion on the accretion-disc lifetime at early evolution stages.\\
We  are searching in stellar Associations for \rm visual triple systems   where all stellar parameters are similar, with the only exceptions of the unknown initial rotation period, and of the circum-stellar environment, in the sense that one of the two about equal-mass components has a close-by third 'perturber' component.\\
In the present study we analyse the 35-Myr old visual triple  system \astrobj{TYC 9300-0891-1}AB + \astrobj{TYC 9300-0529-1} in the young Octans stellar association consisting of three equal-mass K0V components. We collected from the literature all information that allowed us to infer that the three components are actually physically bound forming a triple system and are members of the Octans Association. We collected broad-band photometric timeseries in two observation seasons.   We discovered that all the components are variable, magnetically active,  and \rm  from periodogram analysis  we found the unresolved components \astrobj{TYC 9300-0891-1}AB to have a rotation period P = 1.383\,d and \astrobj{TYC 9300-0529-1} a rotation period P = 1.634\,d. \\
\astrobj{TYC 9300-0891-1}A, \astrobj{TYC 9300-0891-1}B, and \astrobj{TYC 9300-0529-1} have same masses, ages, and initial chemical compositions. The relatively small 16\% rotation period difference measured by us indicates that all components had similar initial rotation periods and disc lifetimes, and the separation of 157\,AU between the component A and the 'perturber' component B (or vice-versa) has been  sufficiently large to prevent any significant perturbation/shortening of the accretion-disc lifetime. 
\end{abstract}
\end{frontmatter}

\section{Introduction}

The early stage of star's life is characterized by the presence of an accretion disc that is magnetically coupled with the central star (see, e.g., \citealt{Menard99}). This coupling implies complex exchanges of angular momentum between the disc and the central star (see, e.g., \citealt{Bouvier07}).
On one hand, the star gains angular momentum from the disc, on the other hand this excess angular momentum must be somehow dissipated. In fact, observations tell us that, until the star-disc interaction is effective, 
the star's rotation period   in most cases \rm remains   about \rm constant (\citealt{Bouvier93}; \citealt{Edwards93}; \citealt{Rebull04}; \citealt{Ingleby14}). This is commonly referred to as star-disc locking. Accretion-driven winds (\citealt{Matt05},  \citealt{Matt08a}, \citealt{Matt08b}) and mass ejection episodes caused by magnetospheric
	reconnection (\citealt{Zanni13}) are among the possible mechanisms that have been proposed to remove the excess angular momentum. Meanwhile, other complex mechanisms, such as hydromagnetic instabilities and intense magnetic fields, also contribute to redistribute internal angular momentum to the star's external envelope (see, e.g., \citealt{Spada10}, \citealt{Spada11}). After the  
 inner disk  dispersion, \rm the stellar radius continues to rapidly decrease while the star is approaching the Zero Age Main Sequence and the rotation rate spins up, despite the angular momentum losses due to magnetized stellar winds (\citealt{Matt07}). The disc lifetime is variable, of the order of a few Myr, but   rarely \rm longer than about 10 Myr (\citealt{Ribas14}; \citealt{Ingleby14}).  A variable disc lifetime implies variable duration of the star-disc locking phase.
\indent
Indeed, the rotation rate of a PMS star at a given age depends on the   accretion \rm disc lifetime, in the sense that
the shorter the disc lifetime the faster the rotation rate with respect to a counterpart star with longer disc lifetime. 
Therefore, the rotation rate can be used to derive information on the disc lifetime.
However, the rotation also depends on the initial rotation period, on the mass and disc mass, and on the environment properties.
We intend to investigate how the environment, and specifically the presence of a nearby companion, can shorten
the disc lifetime and, consequently, make  the rotation spin up to start earlier.
Multiple stellar systems where the components a have equal age, initial chemical composition, and equal mass are 
best suited for this purpose. In these systems, differences between the rotation periods of the components 
are expected to depend only on the initial rotation periods and disc lifetimes.
Specifically, we investigate the hypothesis according to which a close companion can enhance  the 
disc gravitational dispersal allowing one component to start spinning up earlier that the other components. In our first of such cases, the triple system \astrobj{BD$-$21\,1074} in the \astrobj{$\beta$ Pictoris} Association (\citealt{Messina14}), we found that to explain the observed difference of rotation periods between the primary A (P$_{\rm rot} = 9.3$ d) and the secondary B component (P$_{\rm rot} = 5.4$ d), a significant difference of lifetimes of their discs has to be invoked. Specifically, the lifetime of the disc of component B has undergone a shortening owing to enhanced dispersal by dynamical effects from the close-by component C at 15 AU.
Such disc dispersal has shortened the disc-locking phase, allowing component B to start its spinning up earlier than component A.\\
However, the most promising approach to distinguish between different initial rotation rates and different disc locking timescales on a firm basis is a statistical study based on a large number of cases. We are building such a sample by identifying suited stellar systems and deriving the rotation periods of their components.\\
In the present paper, we present the results for another of such systems that has been studied as part of the RACE-OC project (Rotation and ACtivity Evolution in Open Clusters;  \citealt{Messina07}; \citealt{Messina08},  \citealt{Messina10}), while measuring the rotation periods of the members of the about 30--40 Myr \rm old Octans association (\citealt{Messina11}).\\
This is the visual triple system  \astrobj{TYC 9300-0891-1}AB + \astrobj{TYC 9300-0529-1}. This system consists of three equal-masses (0.95\,M$_\odot$), ages ($\sim$35 Myr), and chemical compositions
components physically bound that differ only for their circum-stellar environments and, in principle, for their initial angular momenta. Whereas \astrobj{TYC 9300-0529-1} is quite distant ($\sim$3600\,AU) from \astrobj{TYC 9300-0891-1}AB, on the contrary \astrobj{TYC 9300-0891-1}A has a close-by ($\sim$157\,AU) companion \astrobj{TYC 9300-0891-1}B that may have acted as perturber of its (by now dispersed) accretion disc. Any difference we may find in their rotation periods, should be ascribed to either  the effect of the perturber on the accretion disc, and on its lifetime, or on different initial rotation periods.\\
To measure the rotation periods of \astrobj{TYC 9300-0891-1} and \astrobj{TYC 9300-0529-1} we carried out a dedicated photometric monitoring in two different observation seasons.
In Sect. 2, we present the available information from the literature on these targets. In Sect. 3, we describe our photometric observations. Their analysis is presented in Sect.4, and 5, discussion of the results and conclusions are given in Sect.\,6 and 7. 


\section{The triple system}
 
\astrobj{TYC 9300-0891-1}AB and \astrobj{TYC 9300-0529-1} form a visual triple system, at a mean distance of 174 pc from the Sun,
consisting of three physically bound components having equal mass ($\sim$ 0.95\,M$_\odot$), K0V spectral type, and an age of about 35 Myr.

\astrobj{TYC 9300-0891-1}AB was spatially resolved into a close visual binary by Tycho (\citealt{ESA97}) that
measured an angular separation $\rho$ = 0.9$^{\prime\prime}$, a position angle PA = 128.4$^{\circ}$  at the epoch 1991.25, and an integrated  Johnson V magnitude 
V$_{\rm AB}$ = 10.95$\pm$0.07 mag (\citealt{Hog00})\footnote{
	In the Tycho Double Star Catalogue (TDSC) (\citealt{Fabricius02})
the following magnitudes are reported for both components V$_{\rm A}$ = 11.50$\pm$0.07 
mag and V$_{\rm B}$ = 11.83$\pm$0.11 mag.}. The A and B components are listed as a physical pair with identical proper motions  in the Tycho catalogue. As shown in Sect.\,3, we measured a similar angular separation $\rho$ = 0.95$^{\prime\prime}$ and  similar position angle PA = 135$\pm$10$^{\circ}$. \rm 

\astrobj{TYC 9300-0529-1} is at an angular distance $\rho$ = 20.5$^{\prime\prime}$ from \astrobj{TYC 9300-0891-1}AB (ESA 1997) and 
its Johnson V magnitude (\citealt{Hog00}) is V = 11.75$\pm$0.11 mag.

	In the WDS catalog (The Washington Visual Double Star Catalog, \citealt{Mason01}), \astrobj{TYC 9300-0891-1}AB and \astrobj{TYC 9300-0529-1}
are reported to be a physical pair on the basis of their common proper motions. We have collected from the literature the available kinematic information. Specifically, we retrieved from
	Tycho (ESA 1997), WDS (\citealt{Mason01}), UCAC2 (\citealt{Zacharias04}), and PPMXL (\citealt{Roeeser10}) catalogues the proper motion measurements, and from \citet{Torres06} and \citet{Elliott14} the radial velocity (RV) measurements. Both proper motions and RV measurements of \astrobj{TYC 9300-0891-1}\,AB and \astrobj{TYC 9300-0529-1} are undistinguishable within the uncertainties (see Table\,\ref{param}), indicating that \astrobj{TYC 9300-0891-1}AB and \astrobj{TYC 9300-0529-1} are physically bound and forms a triple system. 

\citet{Torres08} found this system to be member of the young Octans stellar association. However, for no member of this association trigonometric parallaxes exist
and kinematic distances are inferred. Recently, \citet{Murphy14} identified 29 new low-mass members of Octans. Their study allowed to compute accurate mean space motion, the proper motions,  radial velocities and, for the first time, a reliable age estimate of 30--40 Myr based on the Lithium Depletion Boundary. In their study, they inferred kinematic distances of 173 and 175 pc for \astrobj{TYC 9300-0891-1}AB  and \astrobj{TYC 9300-0529-1}. 
We note that the difference between the radial velocities of \astrobj{TYC 9300-0891-1}AB  and \astrobj{TYC 9300-0529-1} is a factor 2.5 smaller that the radial velocity dispersion measured among the Octans
members by \citet{Murphy14}, as well  the difference between the proper motions is a factor 2 smaller that the proper motion
dispersion measured among the Octans (see again  \citealt{Murphy14}). These circumstances further support that the three components are physical bound. \rm
 
The primary A component is a K0V${\rm e}$ emission-line star (\citealt{Torres06}), whereas no spectral classification 
was provided for the secondary B component. However, considering that they have similar V magnitudes (see Sect.\,3), 
we can  infer for the B component 
the same  K0V spectral type.
\astrobj{TYC 9300-0529-1} also has K0V spectral type (\citealt{Torres06}). Therefore, we deal with a system with three
about identical components. We note that \citet{Torres06}
collected only one spectrum of \astrobj{TYC 9300-0891-1}AB (unresolved), which exhibited emission lines, and one of \astrobj{TYC 9300-0529-1}.
However, we know that in the case of young stars, the emission features have variable EW and their detection is rotation-phase dependent (see, e.g. \citealt{Parihar09}). Stars that show emission lines at some epochs may not at other epochs, depending on the activity level and the active region
visibility along the line of sight at the epochs of observation. Additional spectra may reveal emission lines also in \astrobj{TYC 9300-0529-1}.\\



\rm
The system was detected by ROSAT as X-ray source (\astrobj{1RXS\,J184950.1-715703}; \citealt{Voges99}), although the components could not be resolved.
\citet{Torres06}  measured Li EW = 310\,m\AA\,\, for the unresolved component \astrobj{TYC 9300-0891-1}AB and 300\,m\AA\,\, for \astrobj{TYC 9300-0529-1}. 
From these values, \citet{dasilva09} derived the Li abundances  A(Li) = 2.69 and A(Li) = 2.64, adopting T$_{\rm eff}$ = 5168\,K and T$_{\rm eff}$ = 
5140\,K for \astrobj{TYC 9300-0891-1}AB and \astrobj{TYC 9300-0529-1}, respectively.\\
\citet{Torres06}  also measured the projected rotational velocity of \astrobj{TYC 9300-0891-1}AB ($v\sin{i}$ =  9 kms$^{-1}$) and \astrobj{TYC 9300-0529-1} ($v\sin{i}$ = 25.8 kms$^{-1}$). 
 
\begin{figure}
\includegraphics[width=60mm,height=90mm,angle=90, trim= 0 0 0 0]{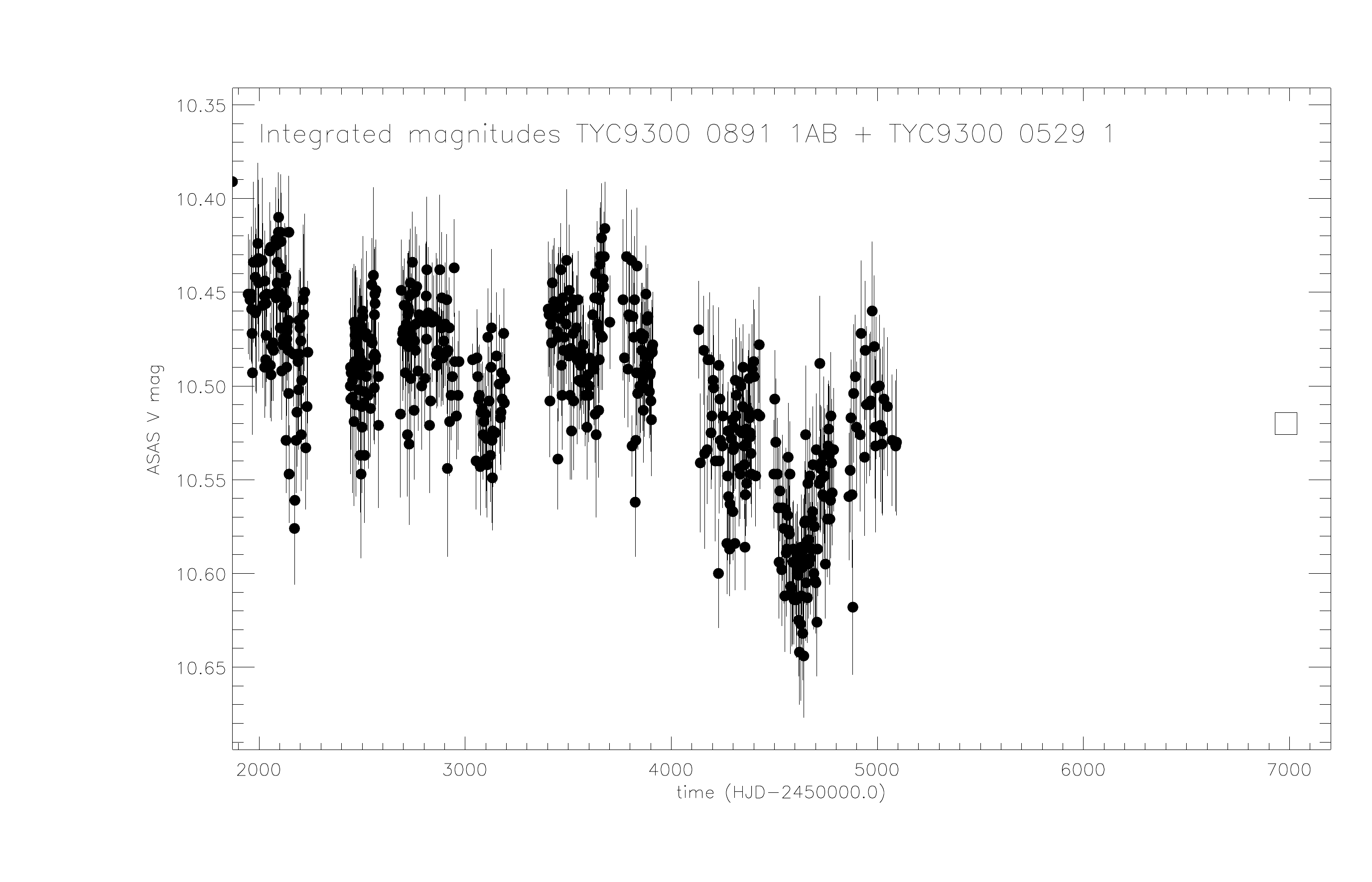}
\caption{V-band magnitude timeseries of the spatially unresolved system TYC9300 0891 1AB + TYC9300 0529 1 as observed by ASAS and at KKO in the 2014 (open square).}
\label{ASAS}
\end{figure}

Since all three components have late spectral type and fast rotation (see Sect.\,4), all are expected to  to be spotted \rm and to exhibit photometric variability induced by such surface inhomogeneities. Actually, our own magnitude measurements differ by up to a few tenths from the Tycho magnitudes measured in the early Nineties.  We could measure the short-term
(rotational) variabilities of \astrobj{TYC 9300-0891-1}AB and of  \astrobj{TYC 9300-0529-1}, which has amplitude up to  $\sim$0.10 mag
in the V band\footnote{Inferred from R-band light curve amplitudes (\citealt{Messina14})}  (see  Sect.\,4). For the long-term variability, we only have integrated magnitudes (owing to the low angular resolution)  of all three components provided by the  All Sky Automated Survey (ASAS; \citealt{Pojmanski97}) (see Fig.\,\ref{ASAS}), which show the effect resulting from the combination of the brightness variability of the three stars. Since all components have similar brightness,  in principle they all can significantly \rm contribute to the observed variability.\\
The Li abundance, X-ray emission, photometric variability, and fast rotation are all consistent with the inferred young age of this system.

\rm


\section{Observations}
This stellar system was observed in 2011 at the South African Astronomical Observatory (SAAO). We could observe it for 6 nights, from April 3 to 12 (see Table\,\ref{obslog}). We used the 1-m telescope 
equipped with STE4 CCD camera and Johnson-Cousins BV(RI)$_c$ filter set. 
We could collect a total of 31 frames in the R band using integration time of 10 sec.
The data reduction was performed using the IRAF\footnote{IRAF is distributed by the National Optical Astronomy Observatory, which 
is operated by the Association of the Universities for Research in Astronomy, inc. (AURA) under 
cooperative agreement with the National Science Foundation.} tasks within DAOPHOT. All frames were bias subtracted and flat field corrected. We used aperture photometry to extract the series of magnitudes of all stars detected in the 5$^\prime$x5$^\prime$ field of view. The two stars \astrobj{2MASS\,J18501087-7156309} (C) and \astrobj{2MASS\,J18495139-7155131} (CK)  (see Table\,\ref{comp}), served as comparison and check, respectively, to get the differential magnitude time series.
The achieved photometric precisions of the differential magnitudes were 0.002 mag and 0.003 mag for 
\astrobj{TYC 9300-0891-1} and \astrobj{TYC 9300-0529-1}, respectively. 

We find the magnitude of the comparison to stay constant during the run with a standard deviation  $\sigma_{C-CK} = 0.013$ mag. The two components A and B of \astrobj{TYC 9300-0891-1} in almost all  nights could not be spatially resolved, and we extracted their integrated magnitudes series.\\
 
\begin{figure}
\includegraphics[width=70mm,height=60mm,angle=0,trim= 0 0 0 0]{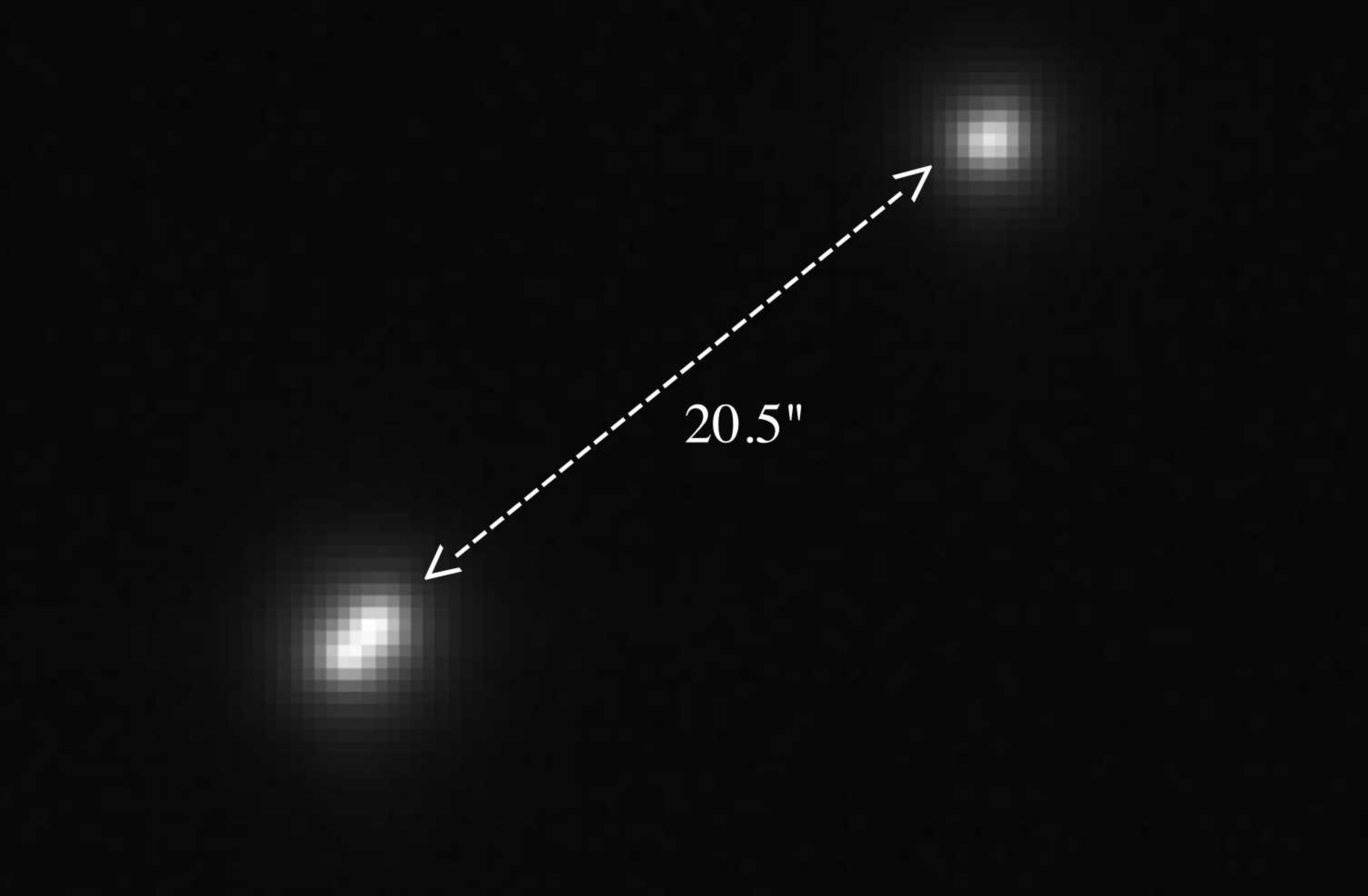}\\
\includegraphics[width=60mm,height=80mm,angle=0,trim= 10 210 10 50]{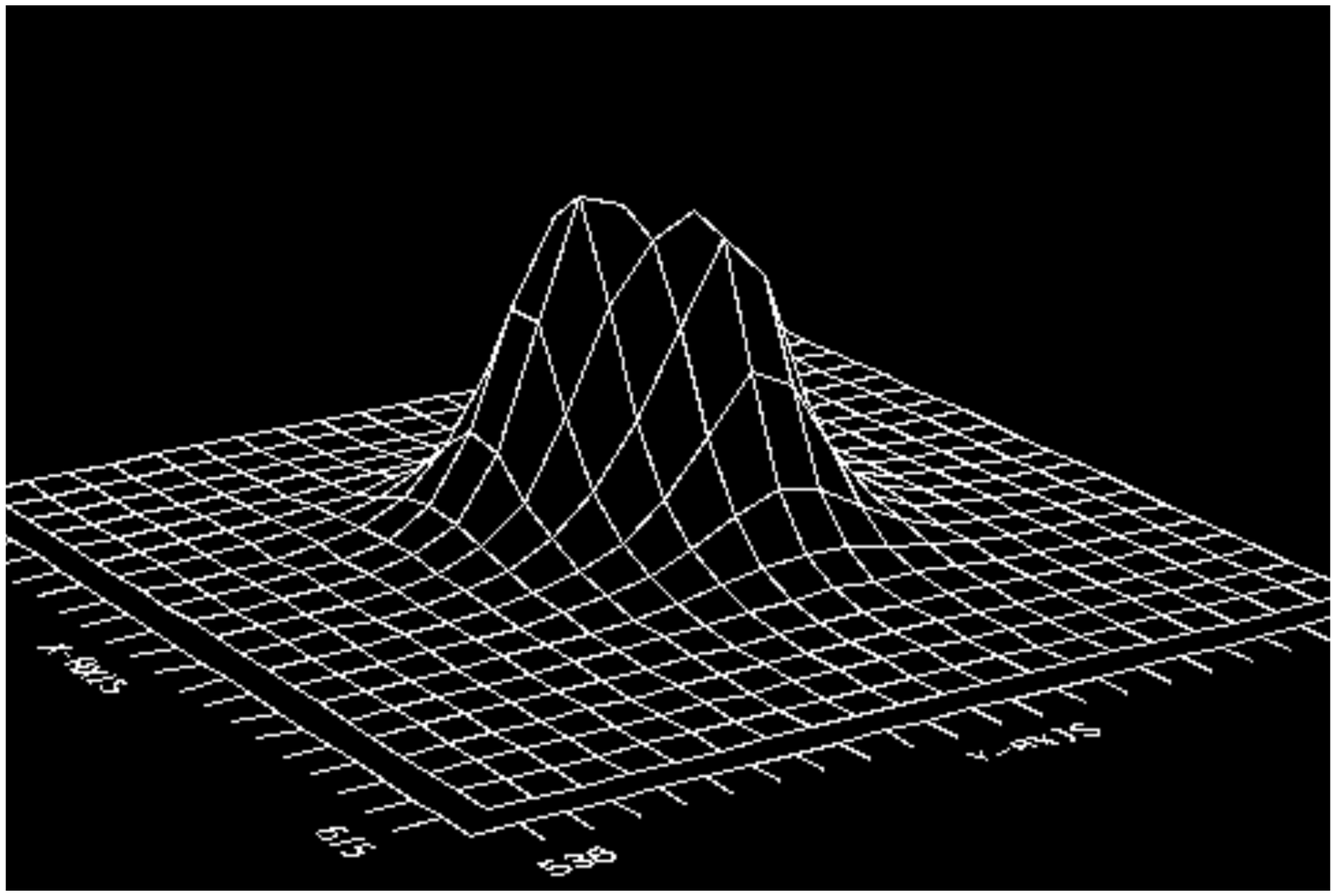}
\caption{\it top panel\rm: section of a frame collected with the 1m SAAO telescope. The components A  and B are at bottom left. \it bottom panel\rm: Surface plot of TYC\,9300-0891-1AB where the two components are partially resolved.}
\label{psf}
\end{figure}

The SAAO STE4 camera mounted on the 1m telescope has a plate scale of 0.31$^{\prime\prime}$/px. On few best-seeing nights, we could resolve partially component A from B, but
sufficiently to extract accurate magnitudes with the PSF technique. In the left panel of Fig.\,\ref{psf}, we show a section of a frame centred on the stellar system and collected on April 10, 2011
where the binary nature of \astrobj{TYC 9300-0891-1} is clearly visible. In the right panel, we plot the surface plot obtained with IRAF which shows the A and B component partially
resolved. The PSF photometry allowed us to extract the following magnitudes V$_{\rm A}$= 11.65$\pm$0.011, V$_{\rm B}$= 11.67$\pm$0.010, and V$_{\rm C}$= 11.68$\pm$0.008 mag.
We measured a separation $\rho$ = 0.95$\pm$0.47$^{\prime\prime}$ between the A and B components, in good agreement with the separation measured by Tycho,  and similar position angle PA = 130$^\circ$$\pm$10 (or PA = 310$^\circ$$\pm$10 depending on which component either A or B is assumed to be the primary). \rm

We carried out a second observing run in 2013 for 45 nights from May 31 until   October 9 (see Table\ref{obslog}) at the Klein Karoo Observatory (225 m a.s.l., Western Cape, South Africa) with a 30-cm (f/8) 
RCX-400 telescope having 21$^\prime\times$14$^\prime$ field of view. It is equipped with SBIG ST8-XME CCD camera and Johnson V and Schuler R filters. 
We could collect a total of 1034 frames in the R band using integration time of 27 sec.   In Nov 21, 2014 we could collect 7 additional frames in the V filter. \rm
On each night, from 4 to 10 consecutive frames were summed up to improve the S/N ratio to get on average from one to three magnitudes per night to be used for the subsequent analysis. 
The data reduction was performed as for the first run data set collected in 2011. Despite the smaller telescope aperture, \astrobj{TYC 9300-0891-1} and \astrobj{TYC 9300-0529-1} 
were spatially well resolved by our observing system, then we used again aperture photometry to extract the series of magnitudes. The best extraction aperture was changed on each night according to variable seeing conditions.
Owing to the larger FoV, we could identify a new set of comparison stars (see Table\,\ref{comp}) with more stable light curves.
The star \astrobj{TYC 9300-0573-1} (C) was used as comparison star to get differential values of our targets.  The achieved photometric precisions of the differential magnitudes were 0.004 mag and 0.005 mag for 
\astrobj{TYC 9300-0891-1} and \astrobj{TYC 9300-0529-1}, respectively. The two stars \astrobj{TYC 9300-1261-1} (CK1) and \astrobj{2MASS\,J18485976-7152445} (CK2) served as check stars. We find the magnitude of the comparison to stay constant during the whole run with  standard deviation  $\sigma_{C-CK1} = 0.009$ and $\sigma_{C-CK2} = 0.011$ mag. 
Of course, the two components A and B of \astrobj{TYC 9300-0891-1} could not be spatially resolved, then we extracted their integrated magnitudes series.\\

\rm
\section{Periodogram analysis}
Our targets, being young low-mass stars, are expected to host some level of magnetic activity and, therefore, are expected to exhibit starspots on their photospheres. The light modulation induced by the star's rotation, which carries these spots in and out of view, is expected to be quasi-periodic and to have the star's rotation period. In these spotted stars, the rotation period can be derived by Fourier analysis of the magnitude time series. In our specific case, to search for the photometric rotation periods we used Lomb-Scargle (LS; \citealt{Scargle82}) and Clean (\citealt{Roberts87}) periodograms.  An estimate of the False Alarm Probability (FAP), that is the probability that a peak of given power in the periodogram is caused by statistical variations, i.e., by Gaussian noise, was done using Monte Carlo simulations according to the approach outlined by \citet{Herbst02}. A more detailed description is given by \citet{Messina10}. The uncertainty on the rotation period determination was estimated following \citet{Lamm04} (see also  
\citealt{Messina10}).

\begin{figure*}
\includegraphics[width=80mm,height=150mm,angle=90]{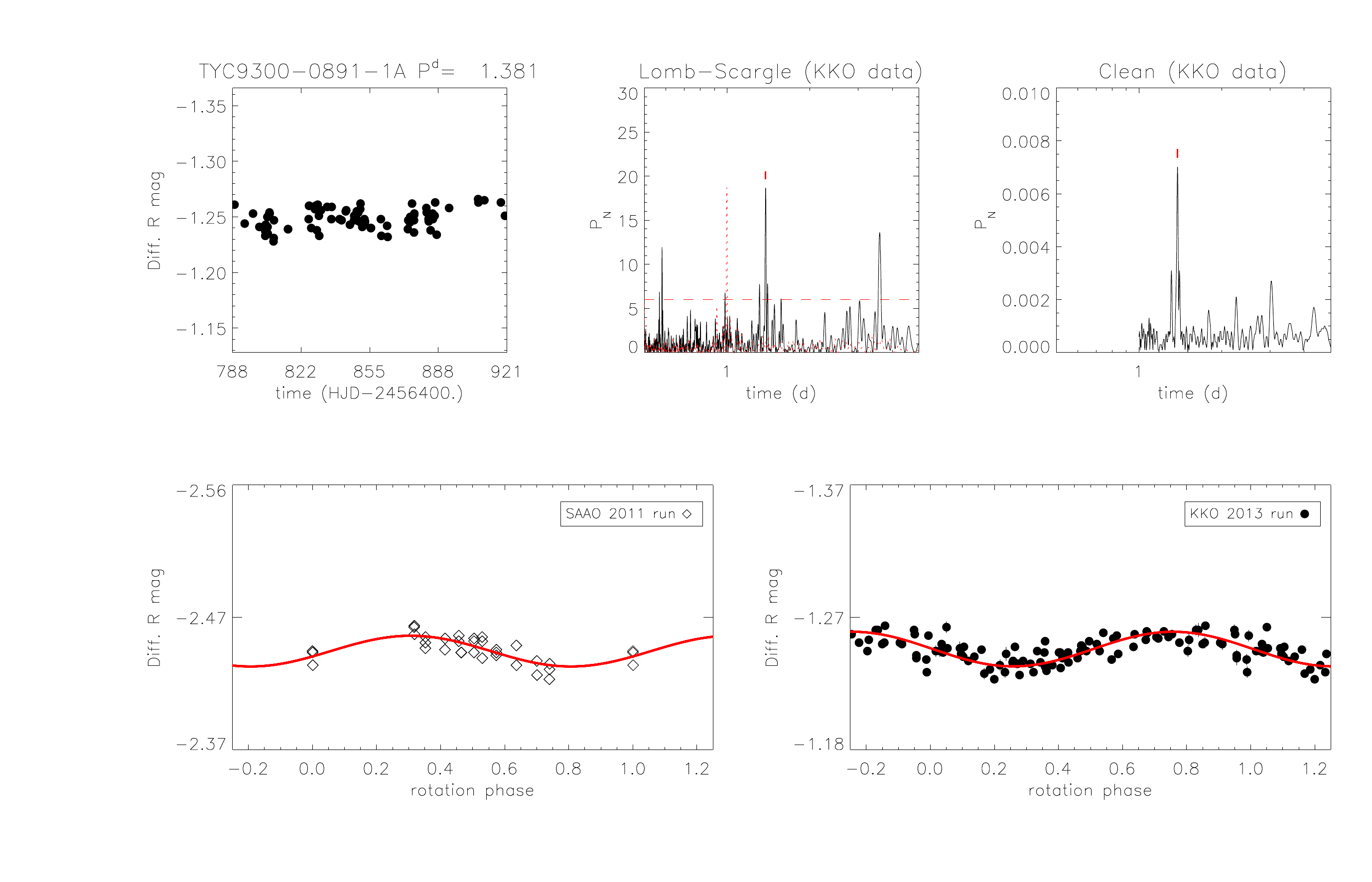}\\
\caption{Top panels: Results of periodogram analysis of TYC 9300-0891-1A. We plot the R-band magnitude timeseries versus Heliocentric Julian Day; the Lomb-Scargle periodogram (solid line) with overplotted the spectral window (dotted red line), whereas the horizontal dashed line represent the 1\% FAP level; the Clean periodogram. Bottom panel: the phase light curve with the rotation period and overplotted the sinusoidal fit (solid red line). 
}
\label{plot}
\end{figure*}

In the case of \astrobj{TYC 9300-0891-1}AB, as already mentioned, the components A and B are unresolved. However, since they have same brightness, in principle both components may significantly contribute to the observed variability leaving signs of their periodicities \rm (that is two power peaks) \rm in the periodogram. The LS periodogram in both seasons revealed  three \rm  significant periodicities   (FAP $<$ 1\% in both seasons): \rm the dominant period is P  = 1.383$\pm$0.001\,d,  
the secondary periods are  P = 3.60$\pm$0.01\,d,  and P = 0.560\,d \rm  (see top of Fig.\,\ref{plot}).  The Clean periodogram showed P  = 1.383\,d, whereas the longest and the shortest  periods were \rm  effectively filtered out by the Clean algorithm. P  = 1.383\,d is the stellar rotation period of either component A or B (more specifically, of the component exhibiting the larger variability), whereas P  = 3.60\,d   and P = 0.560\,d are  beat periods \rm between  P  = 1.383\,d and the data sampling of about 1d, 
imposed by the rotation of the Earth and the fixed longitude of the observation site. The light curve amplitude (amplitude of the fitting sinusoid) is $\Delta$R = 0.02 mag. 
The presence of only one power peak in the cleaned periodogram indicates that  only one component is effectively contributing to the observed variability, the other component having
either no spots at the epochs of our observations or a quite uniform spot distribution in longitude that produces very small rotational modulation, or spots out of view (see Sect. 5.1). \rm
In the top panels of Fig.\,\ref{plot}, we plot the KKO differential magnitude timeseries; the LS periodogram, where the red dotted line is the spectral window and the horizontal dashed line the power corresponding to 99\% confidence level; the Clean periodogram; and the light curve phased with the rotation period with overplotted a sinusoidal fit (diamonds are SAAO data, dots are KKO data).\\

\begin{figure*}
\includegraphics[width=80mm,height=150mm,angle=90]{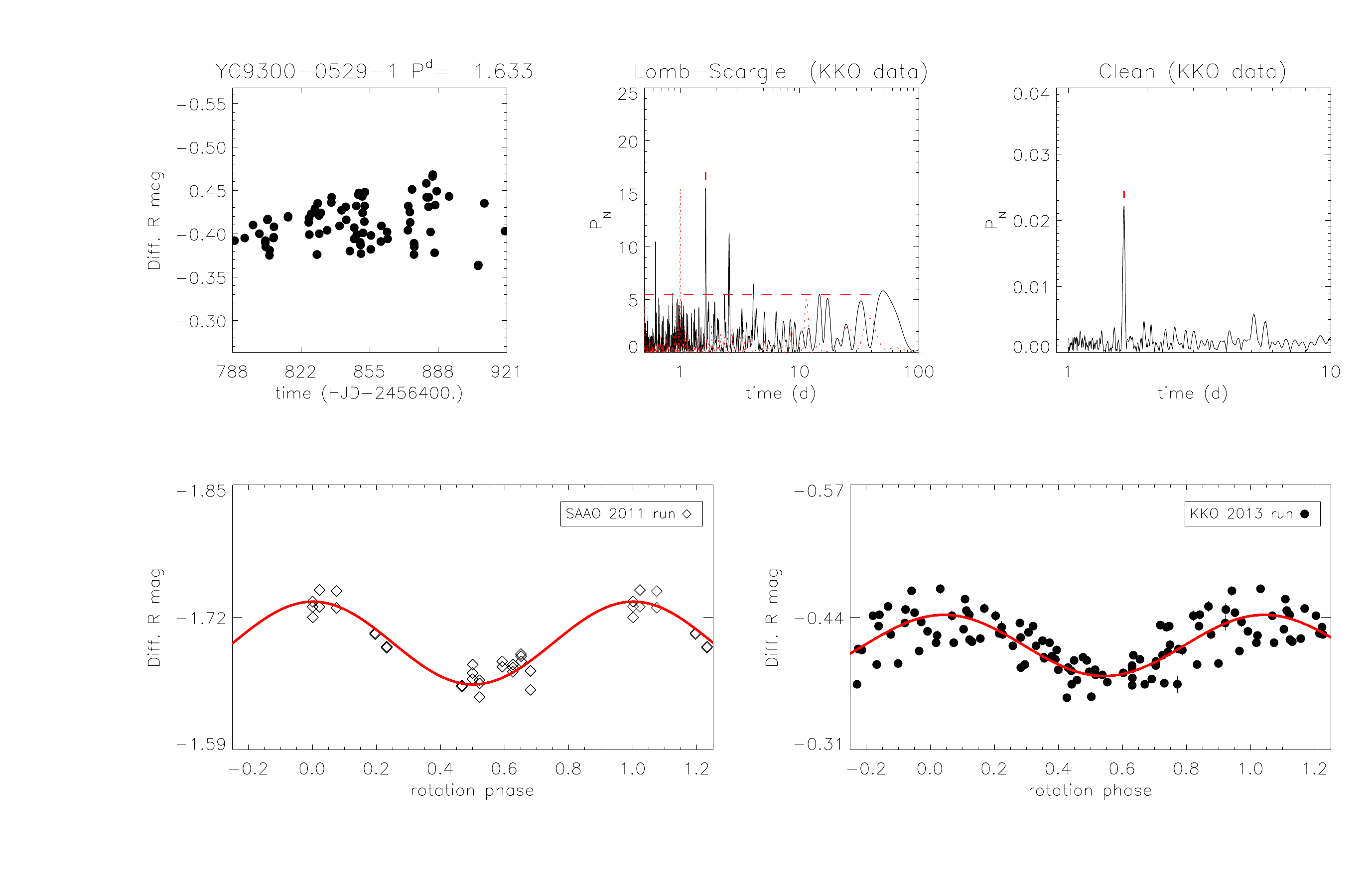}
\caption{Results of periodogram analysis TYC9300-0529-1 (see Fig.\,\ref{plot} for explanation). 
}
\label{plot1}
\end{figure*}

In the case of \astrobj{TYC 9300-0529-1}, our results are summarized in 
 Fig.\,\ref{plot1}. The LS  periodogram in both seasons reveals again   three significant periodicities: \rm the dominant period is P  = 1.634$\pm$0.001\,d, 
and the secondary periods are P  = 2.57$\pm$0.01\,d, which is the same period reported by \citet{Messina11} and derived from the analysis of ASAS timeseries,  and P = 0.62\,d. \rm The Clean periodogram reveals the longer periods in the first run, but only the shorter period P  = 1.632\,d in the second run. All secondary  periods are again alias arising from the spectral window and are effectively filtered out by the Clean algorithm.  We conclude that P  = 1.634\,d is the  stellar rotation period.
The observed light curve has  amplitude $\Delta$R = 0.08 mag in the first season. In the second season, we note that the amplitude has progressively increased from $\Delta$R = 0.03 mag up to  $\Delta$R = 0.06 mag, as as consequence of the active regions evolution.\\

\begin{figure*}
\includegraphics[width=80mm,height=80mm,angle=0, trim= 0 0 0 0]{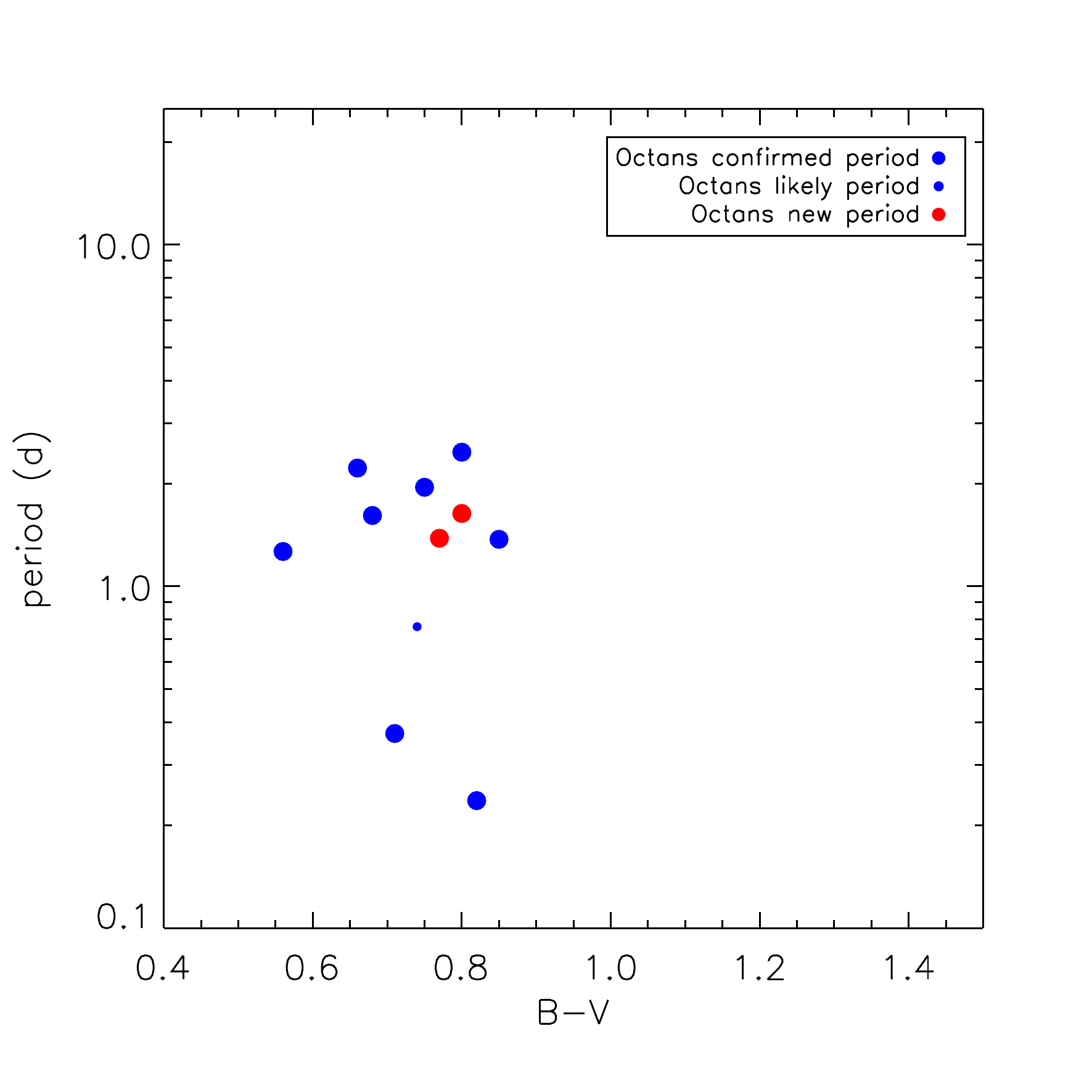}
\includegraphics[width=80mm,height=80mm,angle=0, trim= 0 0 0 0]{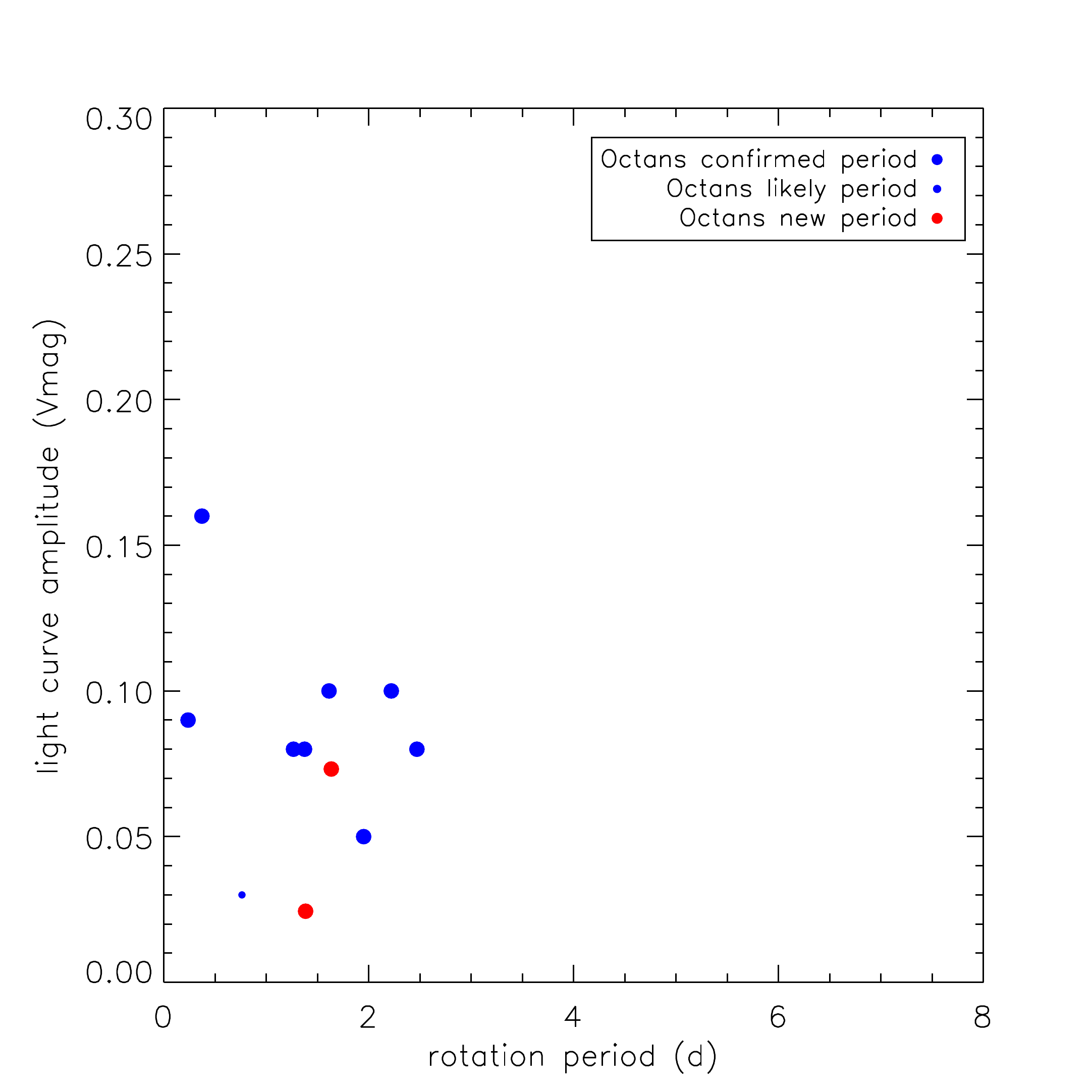}
\caption{Distribution of rotation periods versus B$-$V color (left) and light curve amplitudes versus period (right) of Octans members.}
\label{distri_period}
\end{figure*}

In the left panel of Fig.\,\ref{distri_period}, we plot the distribution of rotation periods versus B$-$V colors of Octans members, which is updated with respect to \citet{Messina11}. We find that our targets have their rotation periods  well within the distribution of the rotation periods of other Octans members in the same color range (0.6 $<$ B$-$V $<$ 0.8). This circumstance further supports the membership
of our target to the Association. \rm 
In the right panel of Fig.\,\ref{distri_period}, we plot the distribution of V-band light curve amplitudes of Octans members versus rotation periods. In our case, the R-band light curve amplitudes of our targets were multiplied by a factor 1.2 to be comparable to the V-band amplitudes (\citealt{Messina14}). \astrobj{TYC 9300-0891-1} is positioned in the lower boundary of the amplitude distribution, owing to the very low inclination of the rotation axis (see Sect.\,5.1).

\section{Target's parameters}

\subsection{\astrobj{TYC 9300-0891-1}A and B}

\begin{figure}
\includegraphics[width=80mm,height=80mm,angle=0, trim= 0 80 0 0]{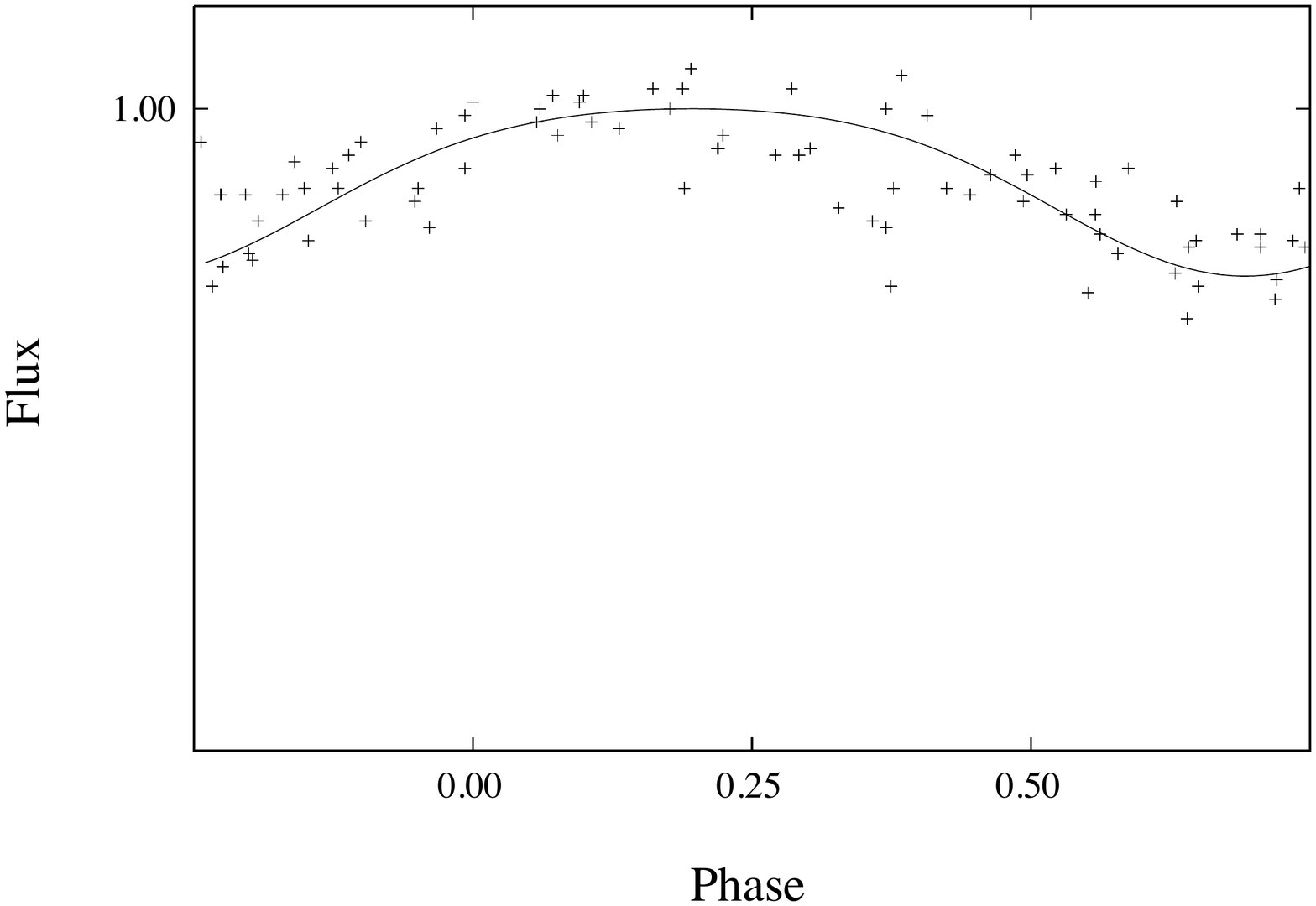}\\
\includegraphics[width=35mm,height=35mm,angle=0, trim= 0 0 20 0]{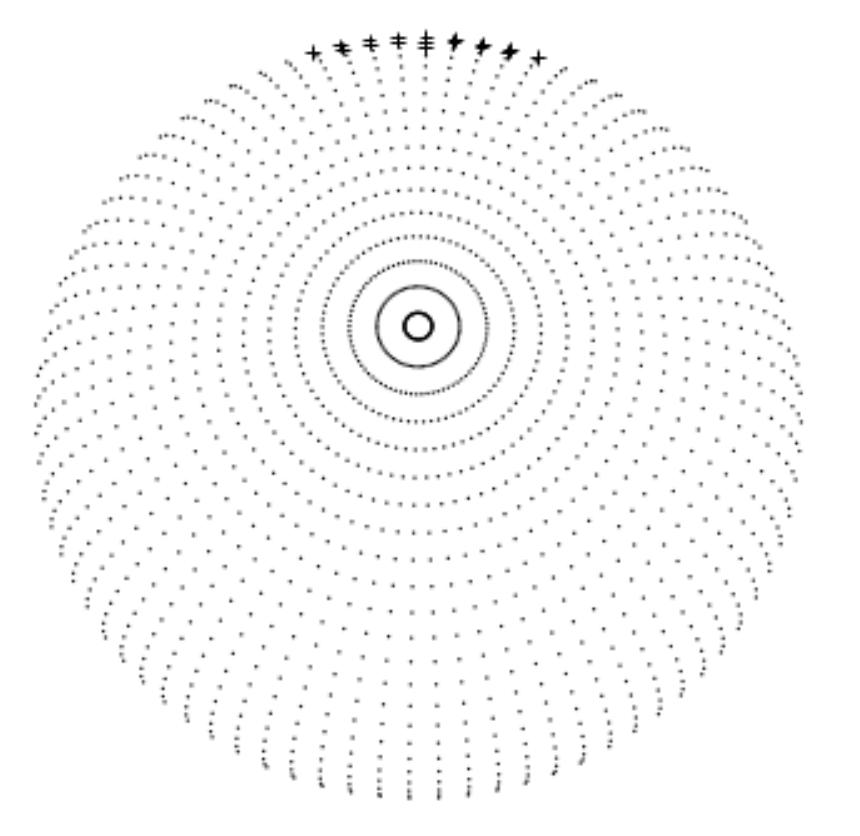}
\includegraphics[width=33mm,height=35mm,angle=0, trim= 0 0 20 0]{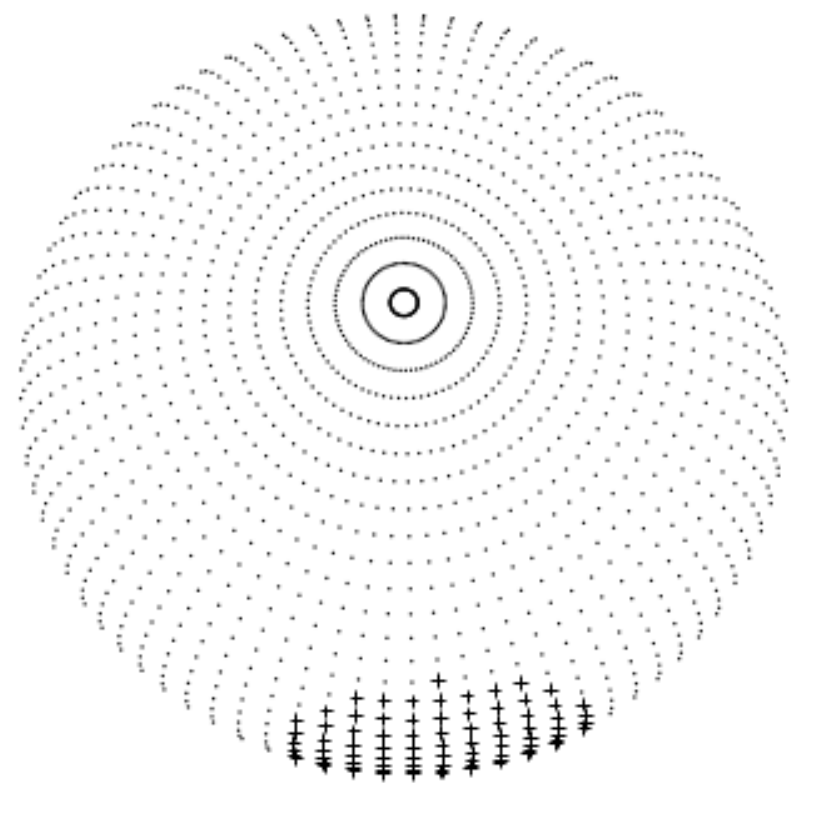}
\caption{Results of spot modeling of TYC 9300-0891-1. \it top: \rm the normalized light curve (crosses) with amplitude $\Delta$F/F = 0.037  phased with the P = 1.383\,d rotation period and  overplotted the model fit (solid line). \it bottom: \rm spot configuration at minimum and maximum spot visibility.}
\label{model_spot}
\end{figure}

\begin{figure*}
\includegraphics[width=80mm,height=150mm,angle=90]{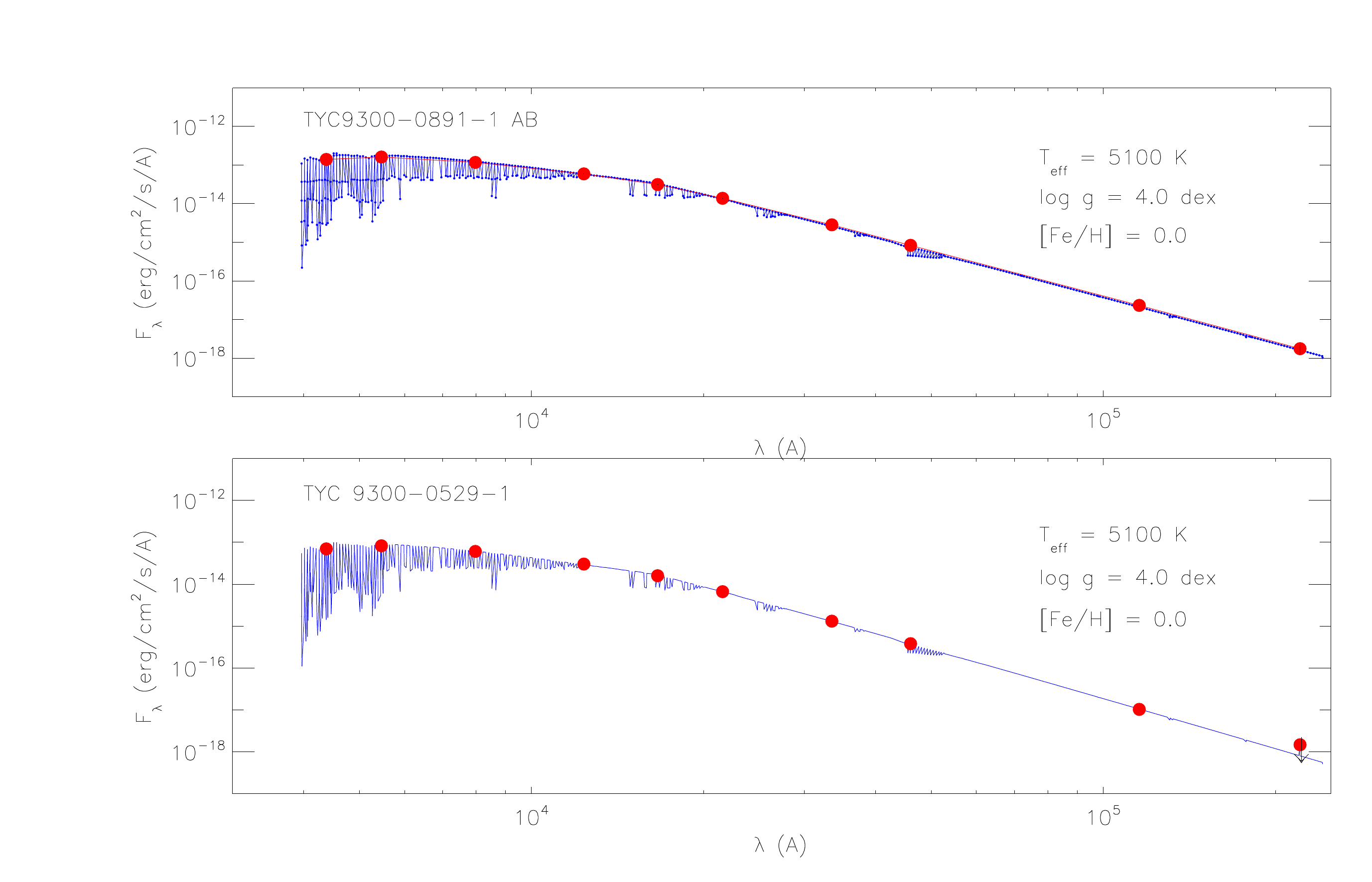}
\caption{Spectral Energy Distributions of TYC\,9300-0891-1AB (\it top\rm) and TYC\,9300-0526-1 (\it bottom\rm). Red bullets are the observed fluxes (see corresponding magnitudes in Table\,\ref{photometry}) whereas blue lines are the best fit synthetic spectra from NexGen Models.  The error associated to the fluxes are smaller than the symbol size.}
\label{sed}
\end{figure*}

We can use effective temperature and luminosity to derive the stellar radius and the inclination of the rotation axis. Assuming d = 173 pc (see Sect.\,2), a mean value of the observed magnitude V = 11.66 mag for both components,  bolometric correction BC = $-$0.33\,mag  (\citealt{Pecaut13}), T$_{\rm eff}$ = 5168\,K (\citealt{dasilva09}), we infer the bolometric magnitude M$_{\rm bol}$ = 5.13$\pm$0.05\,mag, luminosity L = 0.70$\pm$0.06\,L$_\odot$, and the stellar radius R = 1.04$\pm$0.12\,R$_\odot$. Combining the radius with the projected rotational velocity  and the rotation period P = 1.383\,d,  we infer the inclination of the stellar rotation axis  $i$ $\simeq$ 14$^\circ$ for either component A or B. \\
\indent
To check whether \astrobj{TYC 9300-0891-1} could generate detectable flux rotational modulation despite such very inclined rotation axis, we modeled the observed KKO R-band light curve using Binary Maker V2.0 (\citealt{Bradstreet93}). Binary Maker V2.0 models are almost identical to those generated by Wilson--Devinney program (\citealt{Wilson71}) and uses Roche equipotentials to create star surfaces. In our modeling  the second component is essentially ``turned off" (i.e. assigned a near zero mass and luminosity) in order to model a single rotating star (see \citealt{Messina15}).  In Sect.\,4, we showed evidence that only one component has produced the observed variability. \rm  One circular, cool spot of uniform temperature was assumed and the light curve was fit by manual iteration. Gravity-darkening coefficient has been assumed $\nu$ = 0.25 (\citealt{Kopal59}), and  limb-darkening coefficient for the R-band ($u_R$ = 0.60) of \citet{Al-Naimy78} was also adopted. We adopted the effective temperature of the star T$_{\rm eff}$ and inclination $i$ = 14$^\circ$ as input parameters. The temperature contrast between spots and surrounding photosphere  was fixed to T$_{spot}$/T$_{phot}$ = 0.8, which is a value generally measured in K0V spotted stars (see, e.g., \citealt{Berdyugina05}). 
 The 'dilution' effect in the amplitude of the light curve (in our case by a factor 2) arising from the constant flux contribution from the unresolved companion component was also considered (using the 'third light' option
within the model input parameters). \rm
The only free parameters in 
our model remained the spot filling factor and the spot latitude/longitude.
Our aim is not to find unique solution of the spot area and position, but to see if configurations exist that can allow  rotational modulation of 0.02 mag amplitude. In Fig.\,\ref{model_spot}, we plot as an example one of such solutions where in the top we plot the light curve (crosses represent the normalized flux values) with overplotted the model fit, and in the bottom we plot the spot configurations corresponding to the minimum and maximum spot visibility at phases $\phi$ = 0.19 and $\phi$ = 0.69, respectively. In the specific example, we used one circular spot with 25$^\circ$ radius, positioned at the equator.
As we can see, despite the very low inclination value, the star can modulate the spots visibility giving rise to low amplitude flux rotational modulation.\\
  It is likely that at the epochs of our photometric monitoring, one of the two components (either A or B) had either no spots or spots, but out of view owing to the low inclination, thus giving no
contribution to the observed variability and preventing us from the rotation period determination. \rm\\
We use the available optical, near-IR, and IR photometry to build the observed Spectral Energy Distribution (SED). The BVI magnitudes from \citet{Torres06}, JHK magnitudes from 2MASS (\citealt{Cutri03}), and W1-W4 magnitudes from WISE (\citealt{Cutri03}) are listed in Table\,\ref{photometry}. We note that all magnitudes refer to the unresolved AB components.
The SED was fitted with a grid of theoretical spectra from the NextGen Model (\citealt{Allard12}) and the best fit is obtained with a model of T$_{\rm eff}$ = 5100$\pm$100\,K, log g = 4.0$\pm$0.3\,dex and metallicity [Fe/H] = 0.0  (see top panel of Fig.\ref{sed}). We note no evidence of IR excess.

\subsection{TYC\,9300-0529-1}

Assuming d = 175\,pc (see Sect.\,2), brightest observed magnitude V = 11.68 mag,  bolometric correction BC = $-$0.33\,mag  (\citealt{Pecaut13}), T$_{\rm eff}$ = 5140\,K (\citealt{dasilva09}), we infer the bolometric magnitude M$_{\rm bol}$ = 5.13$\pm$0.05\, mag  and luminosity L = 0.69$\pm$0.06\,L$_\odot$, and the stellar radius R = 1.05$\pm$0.11\, R$_\odot$.  From these values we infer the inclination of the stellar rotation axis  $i$ $\simeq$ 50$^\circ$.\\
Although \astrobj{TYC 9300-0529-1} and TYC\,9300-0891-1 have same masses, similar rotation rates, and similar activity levels,   as expected from the empirical rotation-activity relations (\citealt{Messina11}),  \rm  the larger inclination of \astrobj{TYC 9300-0529-1} rotation axis allows a larger amplitude rotational modulation (see Fig.\,\ref{plot1}).\\ \rm
Also for this component we use the available photometry  to build the observed SED (see also Table\,\ref{photometry}). 
The SED was fitted with a grid of theoretical spectra from the NextGen Model (\citealt{Allard12}) and the best fit is obtained with a model of T$_{\rm eff}$ = 5100$\pm$100 K, log g = 4.0$\pm$0.3 dex and metallicity [Fe/H] = 0.0  (see bottom panel of Fig.\ref{sed}).  Also for this star we note no evidence of IR excess. \rm

\begin{figure}
\includegraphics[width=70mm,height=90mm,angle=90, trim= 0 0 0 0]{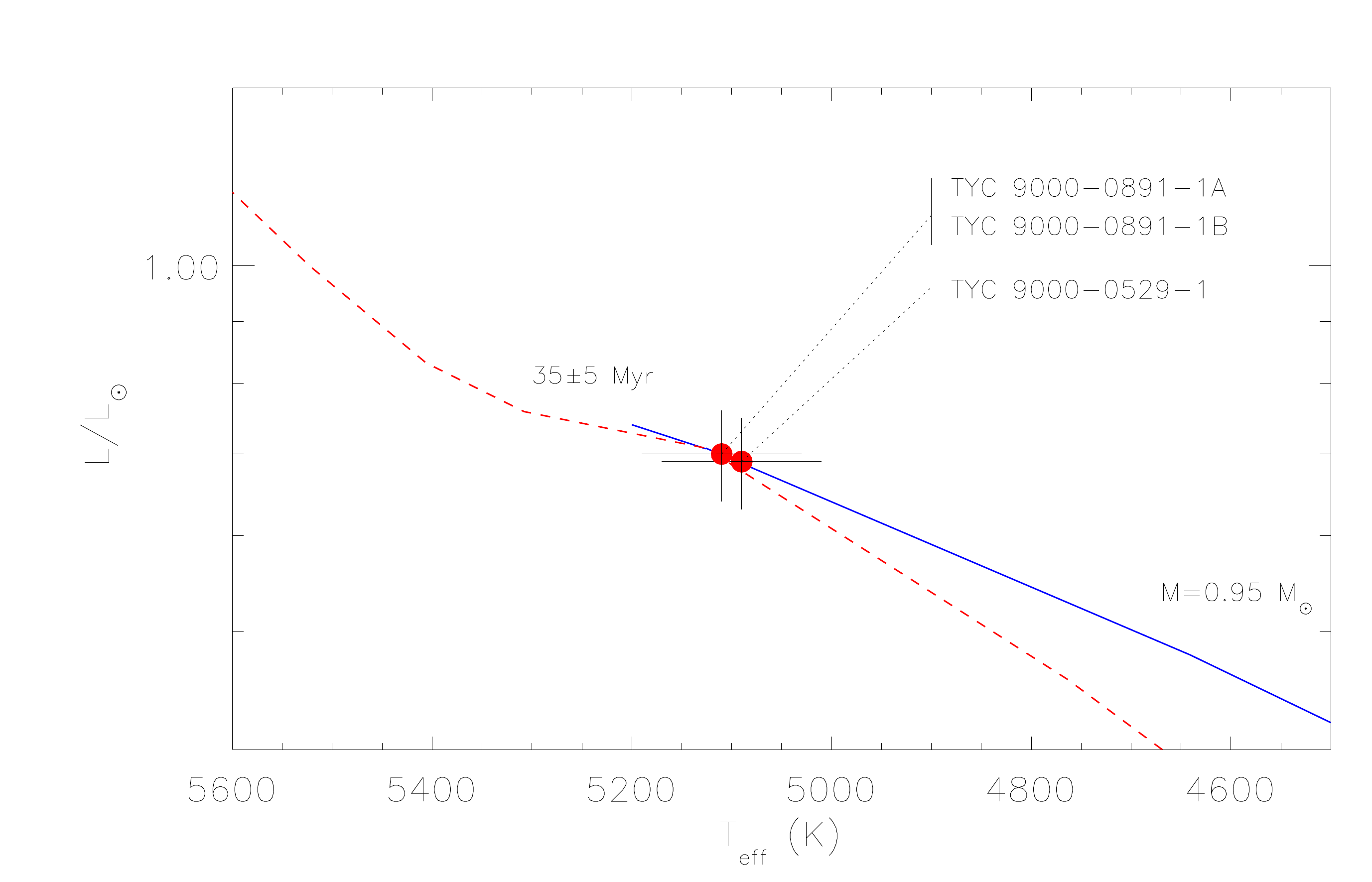}
\caption{HR diagram for TYC\,9300-0891-1 and TYC\,9300-0529-1. The dashed line is the isochrone corresponding to an age of 35 Myr and the solid line is the evolutionary mass track corresponding to 0.95\,M$_\odot$ from  Baraffe et al. (1998).}
\label{hr}
\end{figure}

To infer the mass and put some constrain on the age of the system's components,  we compared T$_{\rm eff}$ and luminosity  with a set of evolutionary tracks and isochronones taken from \citet{Baraffe98} for solar  metallicity.   In Fig.\,\ref{hr}, we see that all three components are fit by an evolutionary  track of M = 0.95$\pm$0.05M$_{\odot}$, and by an isochrone of 35$\pm$5\, Myr, if we take into account the uncertainties. This result further supports the coevalness of all components.\\ \rm

\section{Discussion}

At an age of about 35\,Myr, our targets should have passed the disc-locking phase, and should be experiencing the phase of stellar radius contraction (see, e.g., \citealt{Bouvier14}). Owing to angular momentum conservation, their rotation rates are increasing 
while approaching the Zero-Age Main Sequence. 
 
\begin{figure}
\includegraphics[width=60mm,height=90mm,angle=90, trim= 0 0 0 0]{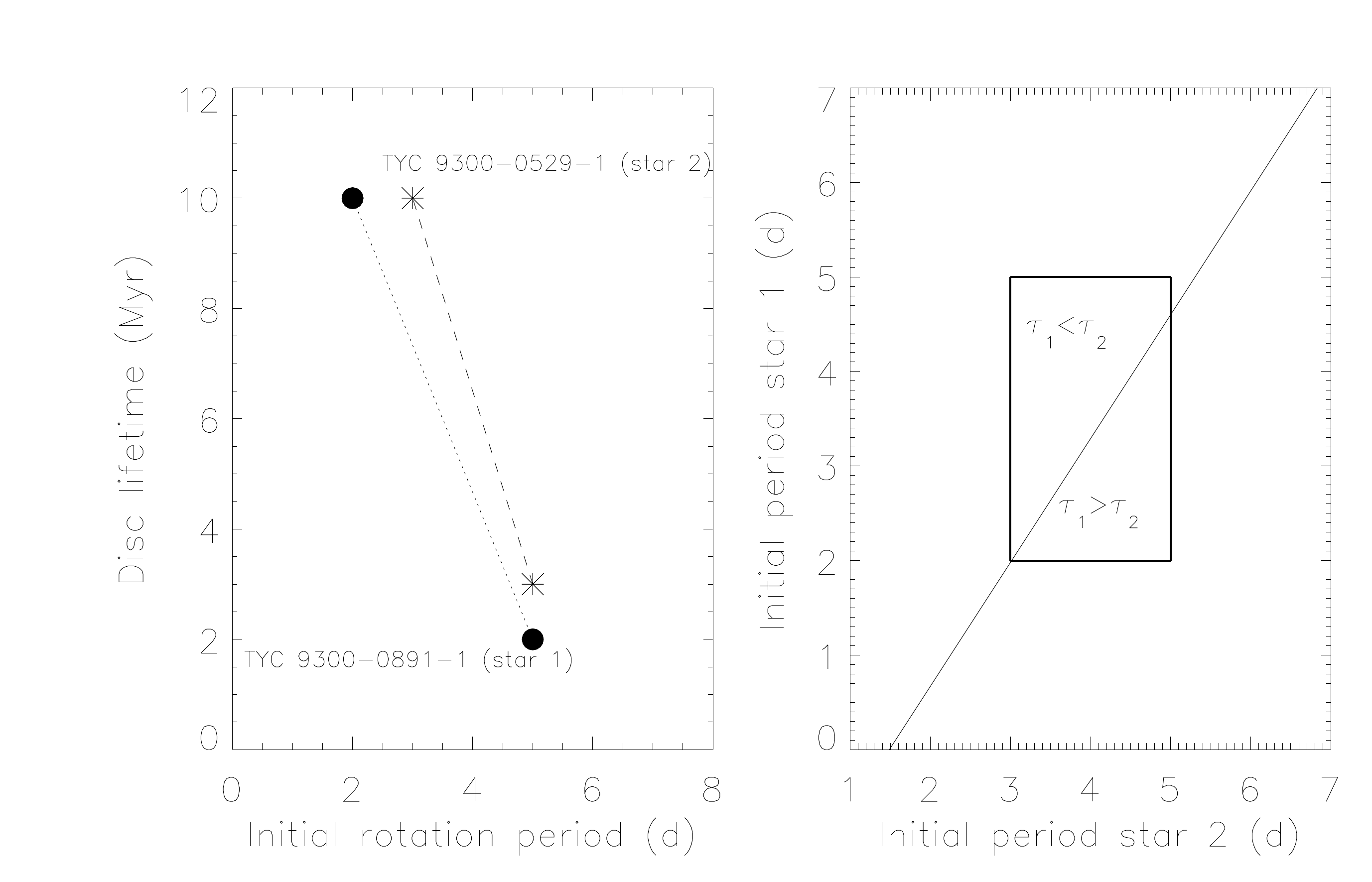}
\caption{\it Left panel\rm: possible values of disc lifetimes versus initial rotation period according to the Gallet \& Bouvier (2013) model for TYC\,9300-0891-1 (star$_1$) and TYC\,9300-0529-1 (star$_2$). \it Right panel\rm: the rectangular area limits the possible combinations of initial rotation periods according to the mentioned model. The currently observed rotation periods can be obtained assuming either  $\tau_1 < \tau_2$, or  $\tau_1 > \tau_2$, or $\tau_1 = \tau_2$. 
}
\label{model_evol}
\end{figure}

Differently than the mentioned case of \astrobj{BD$-$21\,1074}, whose components have rotation periods differing by about 70\%, in the stellar system under analysis the difference amount to only 16\%.
Considering that masses, ages, and initial chemical compositions are almost identical for all components, only disc lifetimes and initial rotation periods may cause this small difference in the rotational evolution of the components. \astrobj{TYC 9300-0891-1}AB is a close binary where the component A has a companion B at a distance of about 157 AU. We want to probe whether this difference in rotation periods may be due to the dynamical effects of component B on the disc of component A (or viceversa) that have facilitated  its dispersal, making shorter the disc lifetime and, consequently, the disc-locking phase. As a consequence,  \astrobj{TYC 9300-0891-1} should have started experiencing the
rotation spin up earlier than the more distant component \astrobj{TYC 9300-0529-1}, determining the observed rotation period difference.\\
\indent
The accretion disc response to an external perturber has been already subject of studies. \citet{Clarke93} analysed the effect of disc dispersal in the case of stellar fly-by. Depending on the kind of encounter, coplanar prograde, coplanar retrograde, and orthogonal, the effect will be more or less destructive on the disc life, the first configuration being the most destructive.  Similar results are found by \citet{Heller95} who computes that less than about half of the initial disc mass is lost during one single encounter, which may be captured by the perturber or ejected by the system. A case more similar to the one under our analysis is discussed by \citet{Korycansky95}. They show that it is possible to model the accretion disc response in an eccentric binary system with succession of parabolic passages. Then, the disc undergoes a series of impacts and looses angular momentum to the secondary, thus reducing its lifetime.\\
\indent
To check our hypothesis, we use as our guideline the angular momentum evolution model of \citet{Gallet13}. We see  in their Fig.\,3 that our targets are mostly similar to the fast rotators  modeled by them. In the age range from about 5 to 30 Myr, the angular velocity linearly increases in  log($\Omega/\Omega_\odot$)-log(age) scale, allowing us to estimate the disc lifetime difference corresponding to the rotation period difference between \astrobj{TYC 9300-0891-1} (hereafter  star$_1$) and \astrobj{TYC 9300-0529-1} (hereafter  star$_2$).\\
However, we first need to make some assumption on the initial values of the rotation periods P$_{in}$ of star$_1$ and  star$_2$ during the disc-locking phase. \\
The rotation period distribution of the Orion Nebula Cluster members at about 1 Myr can be considered as representative of the period distribution during the disc-locking phase. We thus retrieved from \citet{Herbst02} the rotation periods within  a small 
mass range, 0.8--1.2 M$_\odot$, bracketing our components.
The period distribution is found to be double peaked with the shorter periods ranging from about 1 to 5 days. Our first assumption is that our star$_1$ and star$_2$ have their  P$_{in}$ within this range during the disc locking phase. A second assumption is that after 10 Myr the disc-locking in no longer effective and stars are free to spin up. \\
Based on these assumptions, on the present rotation periods, and on the rate of spin up in the 10--30 Myr interval from the Gallet \& Bouvier model, we derive that star$_1$ during the disc-locking phase can have initial rotation period in the range from 5 to 2 days and star$_2$ in the range from 5 to 3 days (see left panel of Fig.\,\ref{model_evol}).
Comparing these two relations we can compute the possible combinations of initial values that will produce the presently observed rotation periods (see rectangular area in the right panel of Fig.\,\ref{model_evol}). More specifically, the solid line represents the combinations when we assume both stars to have same disc lifetimes, whereas the upper region corresponds to the assumption that 
 $\tau_1 < \tau_2$, and the lower region to the assumption $\tau_1 > \tau_2$, where $\tau_1 $ and $\tau_2$ are the disc lifetimes of star$_1$ and star$_2$. The case $\tau_1 < \tau_2$ has a  probability of 57\% with respect to the opposite case. However, on the light of the uncertainties in the comparison with the model, such a difference in probability is not significant and our hypothesis of a lifetime shortening of star$_1$ by dynamical effect by the perturber, although possible, is not favored (more probable). Both scenarios $\tau_1 < \tau_2$ and $\tau_1 > \tau_2$ can account for the observed rotation period difference. \\
 A plausible scenario that comes out from our investigation is that the system's components sheared similar initial rotation periods, and during the 35\,Myr of life have undergone similar rotation period evolution being about similar their disc lifetimes.
Therefore, in this specific system it seems that a stellar companion at a distance of about 157 AU may be still distant enough to produce no effective perturbation on the disc of the other star. The perturbation, if present, but our analysis does not secure us on this issue, may only slightly shorten the disc lifetime to cause a 16\% difference,  at most, between the rotation periods. 
\rm 

On the contrary, in the mentioned visual triple system \astrobj{BD$-$21\,1074} (\citealt{Messina14}) in the 20-Myr $\beta$ Pictoris association, where the component BD$-$21 1074B has a perturber companion BD$-$21\,1074C  at only 15 AU of distance, whereas the component \astrobj{BD$-$21\,1074}A is at 160 AU, the enhanced disc dispersal was effective and allowed the secondary component B to start its spin-up at much earlier stages, reaching a rotation period about 70\% faster than \astrobj{BD$-$21\,1074}A.\\

\section{Conclusions}
We have collected from the literature information that allows us to confidently state that the components \astrobj{TYC 9300-0891-1}AB + \astrobj{TYC 9300-0529-1} are physically bound forming a triple system  and they are member of the Octans association. \rm
We have carried out a photometric monitoring of this system in 2011 and 2013. We could determine the rotation period P = 1.383d for the unresolved components \astrobj{TYC 9300-0891-1}AB and P = 1.641d for \astrobj{TYC 9300-0529-1}. Using archival data, we have modeled their SEDs and  we did not find any evidence for IR excess. \rm We derive the luminosity of all components and comparing with evolutionary model we inferred for the system an age of 35$\pm$5\,Myr, in good agreement with their membership with the Octans Association. Modeling the photometric variability, we found that the photospheric activity level is significant with cool spots covering at least 10\% of the stellar disc.
Since all three components have almost identical mass, age, and initial chemical composition, the rotation period difference likely arises from either different values of the initial rotation periods or different lifetimes of their discs.  However, the 16\% rotation period difference between the components is too small to infer meaningful information on possible differences in the disc evolution history by a comparison with the angular momentum evolution model of \citet{Gallet13}. \\
In the previously examined case of BD$-$21\,1074 in the $\beta$ Pictoris Association,  the period difference was sufficiently large (70\%) to infer that \rm the closer companion at 15 AU determined a more efficient disc dispersal.\\
We are going to study other similar systems to see on a statistical basis how much significant is the role played by such perturber in the primordial disc lifetime or if the observed period distribution is dominated by different values of the initial rotation periods.

\section*{Acknowledgments}  
The Authors thank the anonymous Referees for their helpful comments that allowed to improve the quality of the manuscript. 
 The extensive use of the SIMBAD and ADS databases operated by the CDS centre, Strasbourg, France, is gratefully acknowledged. This publication makes use of VOSA, developed under the Spanish Virtual Observatory project supported from the Spanish MICINN through grant AyA2011-24052.

\newpage

\begin{table}
\caption{\label{obslog}Observation log with listed dates and number of frames in the R filter.}
\begin{tabular}{crcrcr}
\hline
\multicolumn{6}{c}{\bf SAAO}\\
2011/04/03 & 3 & 2011/04/05 & 4 & 2011/04/09 & 6\\
2011/04/04 & 5 & 2011/04/08 & 5 & 2011/04/12 & 8\\
\multicolumn{6}{c}{\bf KKO}\\
2013/05/31 & 8 & 2013/07/12 & 6 & 2013/08/13 & 22\\
2013/06/05 & 12 & 2013/07/15 & 5 & 2013/08/23 & 14\\
2013/06/09& 13 & 2013/07/17 & 14 & 2013/08/24 & 9\\
2013/06/12 & 13 & 2013/07/21 & 6 & 2013/08/25 & 6\\
2013/06/15 & 243 & 2013/07/22 &12 & 2013/08/26 & 17\\
2013/06/16 & 104 & 2013/07/24 & 13 & 2013/09/01 & 21\\
2013/06/17 & 53 & 2013/07/26 & 5 & 2013/09/02 & 14\\
2013/06/19 & 18 & 2013/07/28 & 11 & 2013/09/03 & 6\\
2013/06/20 & 10 & 2013/07/29 & 7 & 2013/09/04 & 17\\
2013/06/26 & 84 & 2013/07/30 & 13 & 2013/09/05 &14\\
2013/07/06 & 19  & 2013/07/31 & 24 & 2013/09/06 &8\\
2013/07/07 & 7 & 2013/08/01 & 28 & 2013/09/12 &6\\
2013/07/08 & 6 & 2013/08/02 & 19 & 2013/09/26 &16\\
2013/07/09 & 8 & 2013/08/03 & 6 & 2013/09/29 &7\\
2013/07/10 & 17 & 2013/08/05 & 21 & 2013/10/07 &6\\
2013/07/11 & 23 & 2013/08/10 & 15 & 2013/10/09 &8\\
                   &      &                    &      & 2014/11/21 & 7 (V)\\
\hline
\end{tabular}
\end{table}

\begin{table*}
\caption{\label{comp} Comparison stars}
\begin{tabular}{llll}
\hline
name & RA(J2000) & DEC(J2000) &  Mag\\
\multicolumn{4}{c}{\bf SAAO}\\
2MASS J18501087-7156309 (C) & 18 50 10.88 & -71 56 30.9 & J = 12.16\\
2MASS J18495139-7155131 (CK) & 18 49 51.40 & -71 55 13.2 & J = 11.82\\

\multicolumn{4}{c}{\bf KKO}\\
TYC 9300-0573-1 (C) & 18 48 29.114 & -71 53 22.48 & V = 11.6\\
TYC 9300-1261-1 (CK1) & 18 48 23.246 & -71 57 15.55 & V = 11.98\\
2MASS\,J18485976-7152445 (CK2) & 18 48 59.763 	& -71 52 44.56 & J = 11.18\\
\hline
\end{tabular}
\end{table*}

 \begin{table*}
 \caption{\label{photometry} Magnitudes from the literature.}

 \begin{tabular}{l@{\hspace{.2cm}}c@{\hspace{.2cm}}c@{\hspace{.2cm}}c@{\hspace{.2cm}}c@{\hspace{.2cm}}c@{\hspace{.2cm}}c@{\hspace{.2cm}}c@{\hspace{.2cm}}c@{\hspace{.2cm}}c@{\hspace{.2cm}}c@{\hspace{.2cm}}}
 \hline
 &   B  & V & I & J & H & K & W1 & W2 & W3 & W4 \\
 &   &   &  &   &  &   &   &   &   &   \\
 \hline
 TYC\,9300-0891-1AB & 11.63 & 10.86 &  9.96 &   9.31 &  8.89 & 8.73 & 8.65 	&	8.67 	&	8.60 	&	8.29\\
 TYC\,9300-0529-1 & 12.39 & 11.59 & 10.68 & 10.04 &  9.63 & 9.52 & 9.48 	&	9.48 	&	9.45 	&	$<$8.93 	\\
 \hline
  \end{tabular}
  \end{table*}

\begin{table}
\caption{\label{param} Summary of information on TYC\,9300-0891-1AB and TYC\,9300-0529-1 from the literature and the present study.}
\begin{tabular}{l c c c l}
\hline
								& \multicolumn{2}{c}{TYC9300-0891-1AB} & \\
								& TYC9300-0891-1A & TYC9300-0891-1B & TYC 9300-0529-1 & \\

								\hline
Sp. Type				& K0Ve	& K0V	& K0V  &  \\
V (mag)					& 11.65 & 11.67  & 11.68   &   this study \\
Mass (M$_\odot$)				& 0.95$\pm$0.05 & 0.95$\pm$0.05 & 0.95$\pm$0.05M & this study\\
T$_{\rm eff}$ (K)				&  \multicolumn{2}{c}{5168$\pm$100} 	&5140$\pm$100 & D09\\
Lum (L$_\odot$) & 0.70$\pm$0.06  & 0.70$\pm$0.06  &0.69$\pm$0.06 &  this study \\
Radius  (R$_\odot$) & 1.04$\pm$0.12 & 1.04$\pm$0.12  & 1.05$\pm$0.12  &  this study\\
log g						& \multicolumn{2}{c}{4.0$\pm$0.3}		& 4.0$\pm$0.3 &  this study	\\
$[$Fe/H$]$					 & \multicolumn{2}{c}{0.0}   & 0.0 &  this study\\
EW Li (m\AA)		& \multicolumn{2}{c}{310}   & 300  & T06 \\
age	(Myr)			& 35$\pm$5      &  35$\pm$5    & 35$\pm$5 &  this study\\
UVW	 (km\,s$^{-1}$)							& \multicolumn{2}{c}{$-$13.7; $-$3.5; $-$11.5} & $-$13.3;$-$4.4;$-$11.4  & M14\\
XYZ		(pc)						& \multicolumn{2}{c}{124.7; $-$93.9; $-$74.7} & 126.1; $-$95.0; $-$75.5 & "  "\\
$\mu_{\alpha}$ mas\,yr$^{-1}$ (Tycho)      & \multicolumn{2}{c}{14.8$\pm$2.1}  & 14.9$\pm$2.1 & ESA (1997)\\
$\mu_{\delta}$ mas\,yr$^{-1}$ (Tycho)      & \multicolumn{2}{c}{$-$16.1$\pm$2.0}        & $-$12.3$\pm$1.9  & "  "\\
$\mu_{\alpha}$ mas\,yr$^{-1}$ (WDS)      & \multicolumn{2}{c}{15}  & 14 & M01\\
$\mu_{\delta}$ mas\,yr$^{-1}$ (WDS)      & \multicolumn{2}{c}{$-$16}        & $-$18 & "   "\\
$\mu_{\alpha}$ mas\,yr$^{-1}$ (UCAC2)      & \multicolumn{2}{c}{14.4$\pm$5.8}  & 14.3$\pm$1.5 & Z04\\
$\mu_{\delta}$ mas\,yr$^{-1}$ (UCAC2)      &\multicolumn{2}{c}{$-$20.4$\pm$5.8}        &$-$18.4$\pm$3. & "  "  \\
$\mu_{\alpha}$ mas\,yr$^{-1}$ (PPMXL)      & \multicolumn{2}{c}{14.8$\pm$2.1}  & 14.1$\pm$1.6 & R10\\
$\mu_{\delta}$ mas\,yr$^{-1}$ (PPMXL      &\multicolumn{2}{c}{$-$16.1$\pm$2.0}        &$-$16.4$\pm$1.6 & "  "  \\

V$_r$ (km\,s$^{-1}$)					& \multicolumn{2}{c}{$-$3.0}		& $-$2.1 & T06 \\
								& \multicolumn{2}{c}{		}	&$-$2.26$\pm$0.47 & E14\\
	distance (pc)		& \multicolumn{2}{c}{173} & 175 & M14\\

\hline
\multicolumn{4}{l}{\small T06: Torres et al. (2006); M14: Murphy et al. (2014); Z04: Zacharias et al. (2004);}\\
\multicolumn{4}{l}{\small R10: Roeser et al. (2010); E14: Elliott et al. (2014); M01: Mason et al. (2001);}\\
\multicolumn{4}{l}{\small D09: da Silva et al. (2009).}\\
\end{tabular}
\end{table}

\newpage

\end{document}